\documentclass[%
reprint,
superscriptaddress,
amsmath,amssymb,
aps,
prl,
floatfix,
]{revtex4-2}
\usepackage{xcolor}

\usepackage{graphicx}% Include figure files
\usepackage{dcolumn}% Align table columns on decimal point
\usepackage{bm}% bold math
\usepackage[colorlinks=true,linkcolor=blue]{hyperref}

\begin{document}
    \title{Exponential Dependence of Interlayer Exchange Coupling in Fe/MgO(001) Superlattices on Temperature}
	\author{Nanny Strandqvist}
	\author{Tobias Warnatz}
	\affiliation{Department of Physics and Astronomy, Uppsala University, Box 516, SE-75120 Uppsala, Sweden}
	\author{Kristbjörg Anna Thórarinsdóttir}
    \affiliation{Science Institute, University of Iceland, Dunhaga 3, IS-107 Reykjavik, Iceland}
    \author{Alexei Vorobiev}
    	\affiliation{Department of Physics and Astronomy, Uppsala University, Box 516, SE-75120 Uppsala, Sweden}
    \affiliation{Institute Laue-Langevin, 71 avenue des Martyrs, 38042 Grenoble, France}
    \author{Vassilios Kapaklis}
	\author{Björgvin Hjörvarsson}
	\affiliation{Department of Physics and Astronomy, Uppsala University, Box 516, SE-75120 Uppsala, Sweden}

\begin{abstract}
    The interlayer exchange coupling in Fe/MgO(001) superlattices is found to increase exponentially with decreasing temperature. Around 150~K, the field induced response changes from discrete switching—governed by field-driven domain propagation—to a collective rotation of the magnetic layers. This transition is accompanied by a change in the magnetic ground state from 180$^{\circ}$ (antiferromagnetic) to 90$^{\circ}$ alignment between adjacent Fe layers. These effects are argued to arise from quantum well states, defined by the total thickness of the samples.
\end{abstract}

\maketitle

Artificial superlattices are fascinating materials that provide a versatile platform for exploring emergent physical phenomena. Furthermore, their properties can be tailored by \emph{e.g.} the choice of constituents, the thickness, and the thickness ratio of the layers  \cite{Esaki_1973, Esaki_1977, GOSSARD19793, Schuller_PRL_1980}. In semiconductor superlattices, the proximity of layers with different band gaps leads to the formation of quantum well states, with their own electronic structure and properties \cite{Esaki_1970}. Similar quantum effects are observed in metallic superlattices where magnetic layers couple across non-magnetic layers through oscillatory interlayer exchange coupling (IEC) \cite{bruno_theory_1995}. In layers, such as MgO, the coupling strength is known to decrease exponentially with increasing thickness of the layer, which is typically attributed to spin-polarized tunneling \cite{faure-vincent_interlayer_2002, chiang_oxidation-induced_2009, koziol-rachwal_antiferromagnetic_2014, Moubah2016}.

Almost all prior studies on IEC in Fe/MgO heterostructures are based on trilayers, with a primary focus on device-related functionalities \cite{parkin_giant_2004, yuasa_giant_2004, bhatti_spintronics_2017, lou_demonstration_2008, parkin_magnetic_2008}. It is only recently the investigation of interlayer exchange coupling has been extended to include Fe/MgO(001) superlattices \cite{Moubah2016, Magnus2018, warnatz_impact_2021, Anna_Fe_Thickness}. These exhibit discontinuous steps in the field response (digital hysteresis), arising from switching of individual Fe layers—a behavior attributed to the interplay between antiferromagnetic IEC and the intrinsic magnetic anisotropy of the Fe layers \cite{Moubah2016}. Moreover, the magnetic response has been found to depend on the number of Fe/MgO repetitions, which was argued to arise from long-range magnetic IEC \cite{Magnus2018, warnatz_impact_2021}. The thickness of the Fe layers also has a profound effect on the strength of the IEC \cite{Anna_Fe_Thickness}. These findings highlight the need to account for all length-scales in the structure—including the MgO spacer, Fe layer, and the total thickness of the stack—when describing the coupling. In this Letter, we present the temperature dependence of the coupling and consider the underlying principles behind the obtained IEC in Fe/MgO(001) superlattices.

    \begin{figure}[t]
        \centering
        \includegraphics[width=0.95\linewidth]{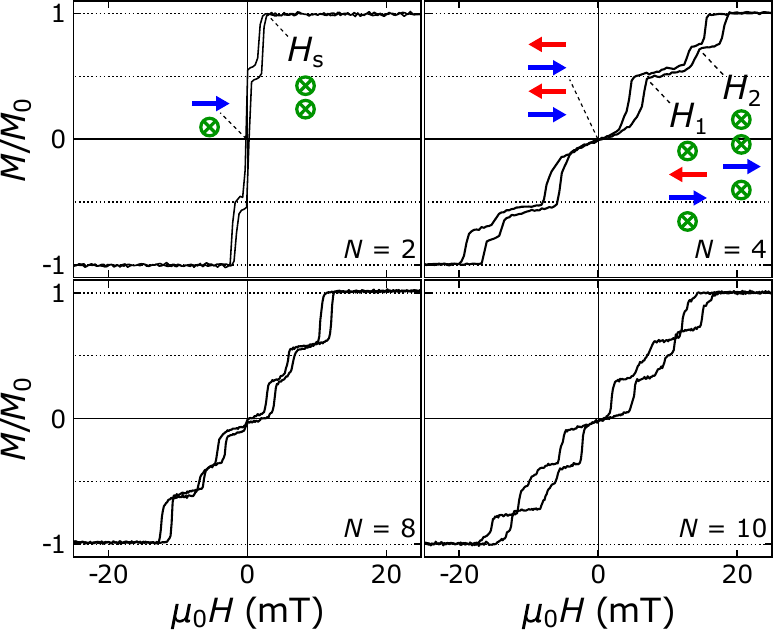}
        \caption{\label{fig:fig1} In-plane hysteresis measured at room temperature by MOKE, along the Fe[100] easy axis for [Fe(2.0\,nm)/MgO(1.7\,nm)]$_{N}$(001) superlattices with $2 \leq N \leq 10$. The labels $H_{\mathrm{1}}$, $H_{\mathrm{2}}$, and $H_{\mathrm{s}}$ denote the switching fields associated with each magnetization step, where $H_{\mathrm{s}}$ corresponds to the field required to align all Fe layers.}
    \end{figure}

The growth and the structural characterization of the superlattices is briefly described in the supplemental material and cited references. 
The magnetic response was investigated using the magneto-optical Kerr effect (MOKE) and Vibrating Sample-Magnetometer (VSM). Normalized hysteresis loops for [Fe(2nm)/MgO(1.7nm)]$_{N}$(001) superlattices with $N =$~2,~4,~8 and 10 bilayer repetitions are shown in Fig.~\ref{fig:fig1}. These measurements were conducted at room temperature with an external magnetic field applied along a Fe~[001] easy axis, and the hysteresis loops were corrected by removing the coercivity of the Fe layers, as described in \cite{warnatz_impact_2021}, to better emphasize the reversibility—or lack thereof—in the switching behavior of the Fe layers. All samples exhibit discrete magnetic switching characterized by step-like features in the magnetization curves. These steps arise from the nucleation and propagation of $90^\circ$ domain walls across the full lateral extent of the 2~nm-thick Fe layers across the sample area of 1~cm$^{2}$ \cite{Chung2013, Magnus2018, Xu2018}. The switching is driven by the applied magnetic field, while IEC and magnetocrystalline anisotropy act as restoring forces. Consequently, the field value at each step is proportional to the strength of the IEC acting on the switching layer, when the field is aligned with the easy axis.

For $N = 2$ (see Fig.~\ref{fig:fig1}a), a magnetic field of approximately 2.4~mT is required to align the magnetic moments of both Fe layers ($H_{\mathrm{s}}$). Upon reducing the field from saturation to zero, a single switching event is observed, resulting in a remanent magnetization of $M/M_{0} \approx 0.5$. This value is consistent with the reversal of a single Fe layer, leading to a 90$^{\circ}$ alignment of the layers. The configuration is likely to be metastable, arising from the competition between a weak antiferromagnetic interlayer exchange coupling and the significantly stronger four-fold magnetocrystalline anisotropy \cite{Bellouard2008, Moubah2016, Magnus2018, Warnatz2020}.

    \begin{figure}[t]
		\centering
		\includegraphics[width=1\linewidth]{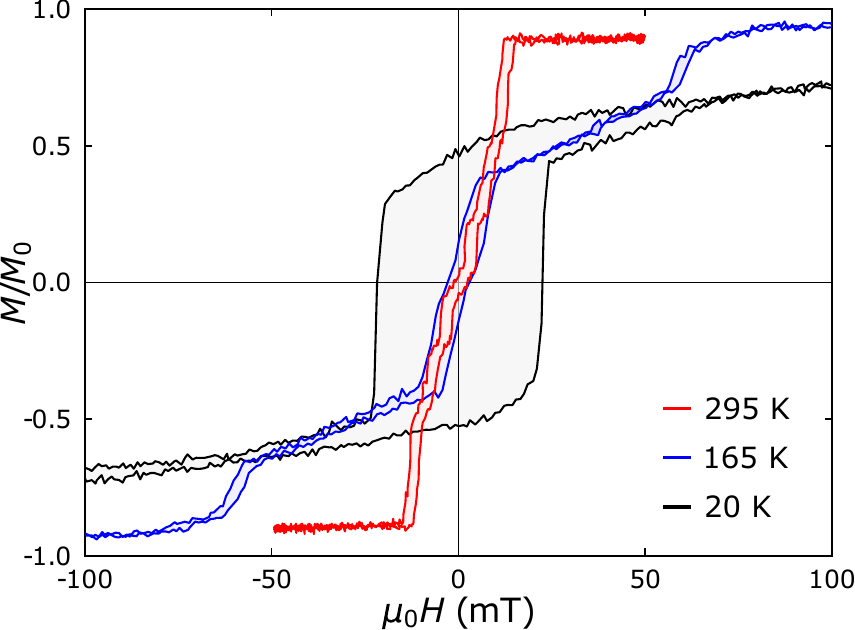}
		\caption{\label{fig:fig2}Normalized hysteresis loops measured along the Fe [100] easy direction for [Fe(2nm)/MgO(1.7nm)]$_{10}$ superlattice at 295, 165, and 20K (MOKE). The moment of the Fe layers is only modestly affected while the saturation field and the coercivity exhibits much stronger temperature dependence.}
	\end{figure}
    
Increasing the number of bilayers to four ($N = 4$) has a pronounced effect on the hysteresis curve, as shown in Fig.~\ref{fig:fig1}b. In particular, the remanence at zero field becomes negligible, consistent with an antiferromagnetic alignment of the Fe layers. Three discrete steps are observed between remanence and saturation. The first step, occurring at $H_{\mathrm{1}}$, is approximately twice as large as the subsequent steps, corresponding to a magnetization change of $M/M_{\mathrm{0}} \approx 0.5$. This step is attributed to the simultaneous switching of the top and bottom Fe layers, which each have only one nearest neighbor and therefore require a lower switching field—an effect that arises purely from the differing number of magnetic neighbors \cite{warnatz_impact_2021, Magnus2018}. As the external field increases further, the remaining Fe layers are sequentially switched. The field required to align all four Fe layers parallel to the external field is $H_{\mathrm{s}} \approx 16$~mT. Similar behavior is observed for $N = 8$ and $10$, as shown in Fig.~\ref{fig:fig1}c–d. A clear change in the relative strength of the coupling is therefore observed. For example, in the sample with $N = 10$, the saturation field is approximately three times larger than $H_{\mathrm{1}}$. In the case of nearest neighbor interactions, $H_{\mathrm{s}}$ would be expected to be twice as large as $H_{\mathrm{1}}$. Consequently, rationalization of the switching, within and between samples, requires interactions beyond nearest neighbor.

Having established the switching properties of the samples at room temperature, we now examine the temperature dependence of the coupling strength in a sample with 10 repeats. For these measurements, the sample was initially cooled to 20~K, and in-plane hysteresis loops were recorded at temperature intervals of 5~K up to 300~K. As seen in Fig.~\ref{fig:fig2}, the shape of the hysteresis curve changes markedly with temperature, while the overall Fe moment is only modestly affected. Two discrete switching steps are observed at $T = 165$~K and the magnetization varies linearly between them. In addition, the field required to align the Fe layers parallel to the external field ($H_{\mathrm{s}}$) increases to approximately 60~mT, indicating a substantial enhancement of the interlayer exchange coupling compared to room temperature. At $T = 20$~K, the (minor-) hysteresis loop is S-shaped, bearing little resemblance to the room-temperature response. A large coercivity is observed, and a remanent magnetization of 0.5$M_{\mathrm{s}}$. At this temperature, the discrete steps associated with digital hysteresis have vanished and the field response is consistent with a coherent rotation of the layers.

      \begin{figure}[t!]
        \includegraphics[width=0.95\linewidth]{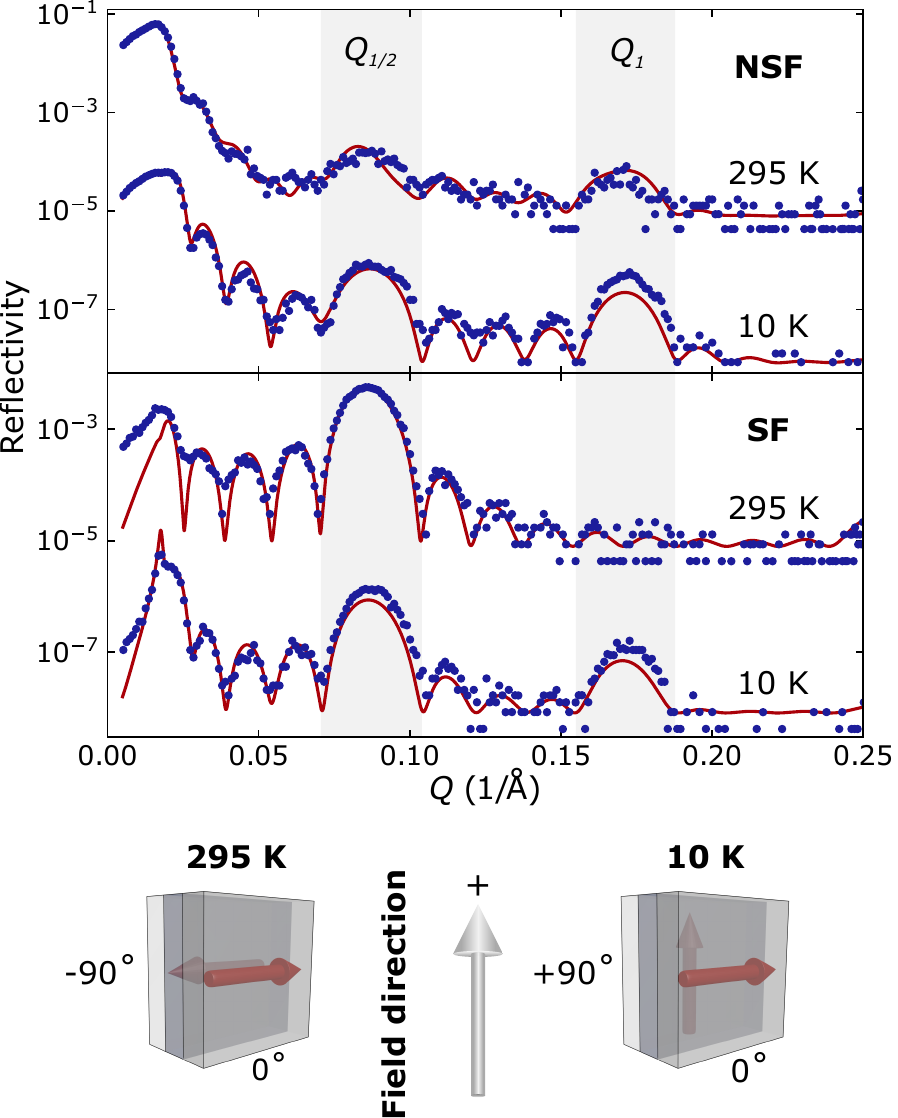}
        \caption{\label{fig:fig3}PNR measurements at 295 and 10~K. Polarized neutron reflectivity data collected in applied fields near remanence (1.5~mT at 295~K and 20~mT at 10~K). The upper panel shows the non-spin-flip (NSF, $R^{++}$) channel, and the lower panel shows the spin-flip (SF, $R^{-+}$) channel. Blue dots represent the experimental data, while red curves are the corresponding fits. Shaded gray regions mark the position of the $Q_1$ and $Q_{1/2}$ peaks.}
    \end{figure}
    
To obtain better understanding of the temperature dependent changes in the switching of the layers, we employed polarized neutron reflectivity (PNR). The PNR measurements were performed on the SuperADAM reflectometer at the Institut Laue-Langevin in Grenoble, France \cite{Vorobiev2015}. The neutron wavelength was set to 5.21~Å, with polarization and analyzer efficiencies of 99.8 and 99.9\%, respectively, at the incident and receiving ends. The sample was cooled using a liquid helium cryostat. Data reduction was carried out using the \texttt{pySAred} software package \cite{Klechikov2021}, which compensates for fluctuations in neutron flux through normalization to integrated monitor counts. The PNR data were fitted to extract the magnetic profile, with structural parameters (see supplementary material)  held fixed and only the magnetization angles within the Fe layers allowed to change. Structural parameters were obtained by simultaneous fitting of XRR and high-field (500~mT, saturated state, at room temperature) PNR data using the \texttt{GenX} software package \cite{Bjorck2007, Glavic:ge5118}.

      \begin{figure}[t]
        \centering
        \includegraphics[width=0.95\linewidth]{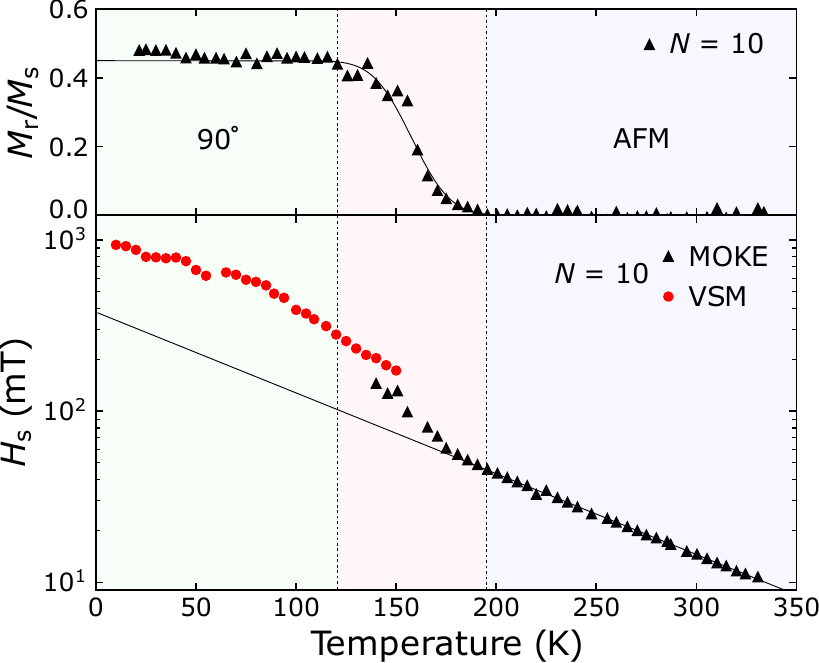}
        \caption{\label{fig:fig4}Temperature dependence of magnetic response. (Top) Remanent magnetization $M_{\mathrm{r}}$ of the [Fe/MgO]$_{10}$ superlattice as a function of temperature, normalized to the saturation magnetization $M_{\mathrm{s}}$. A transition in magnetic ordering is observed around 150~K. (Bottom) Saturation field $H_{\mathrm{s}}$ as a function of temperature. Exponential change in $H_{\mathrm{s}}$ is observed over a broad temperature range. VSM was used for measurement below 165 K.}
    \end{figure}
    
Prior to each measurement, a magnetic field of 500~mT was applied in-plane along the easy [001] axis, after which the field was reduced to the desired value. The magnetic configuration of the layers was extracted by fitting the non-spin-flip (NSF) reflectivity channels $R^{++}$ and $R^{--}$, along with the spin-flip (SF) channel $R^{-+}$. Fig.~\ref{fig:fig3} shows the NSF ($R^{++}$) and SF ($R^{-+}$) PNR traces at 295~K and 10~K for the sample with $N = 10$. The applied fields were close to remanence: 1.5~mT at 295~K and 20~mT at 10~K. At 295~K, the NSF channel displays a first-order Bragg peak at a scattering vector $Q_1 = \frac{2\pi}{\Lambda} = 0.172$~Å$^{-1}$, originating from the structural periodicity of the superlattice, where $\Lambda$ is the Fe/MgO bilayer thickness. In addition, a weak half-order peak $Q_{1/2}$ is observed, with its presence indicating a magnetization component along the applied field direction with twice the structural periodicity. The SF channel, which is sensitive to transverse magnetic components, shows a well-defined $Q_{1/2}$ peak, indicating alternating transverse components in every second Fe layer. Simultaneous fitting of the PNR and XRR data yields an average angle of 180$^{\circ}$ between adjacent Fe layers. The layers are therefore antiferromagnetically ordered at room temperature. At 10~K, the PNR data, see  Fig.~\ref{fig:fig3}, exhibit well-defined $Q_1$ and $Q_{1/2}$ peaks in both the NSF and SF channels. These features arise from alternating parallel and perpendicular magnetization components in every second Fe layer. Fitting the data reveals that every second layer is aligned with the applied field, while the intervening layers are oriented nearly perpendicular, as illustrated in Fig.~\ref{fig:fig3}.

   \begin{figure}[t]
		\centering
		\includegraphics[width=0.95\linewidth]{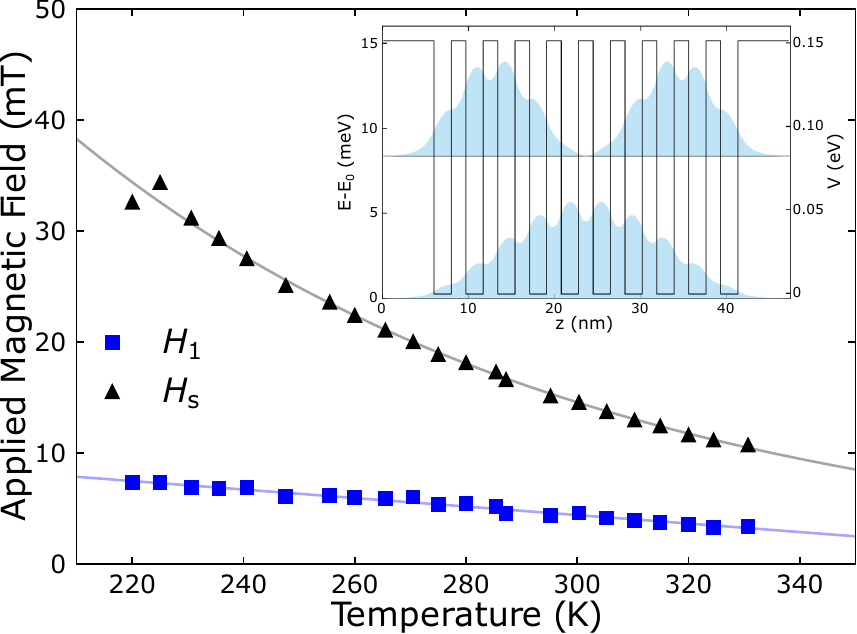}
		\caption{\label{fig:fig5}Temperature dependence of the field required to switch the outermost Fe layers ($H_{\mathrm{1}}$) in a sample with $N = 10$ and to align all layers parallel to the external field ($H_{\mathrm{s}}$) along the easy axis ([001]). The field $H_{\mathrm{1}}$ increases linearly with decreasing temperature and is fitted by $H_{\mathrm{1}}(T) = a\cdot T + b$ ($a=-0.04$, $b=16$), while $H_{\mathrm{s}}$ is fitted using $H_{\mathrm{s}}(T)= a\cdot e^{-b T}$ ($a=365$, $b=0.99$). The inset shows calculated collective quantum well state formed in a [Fe(2.0\,nm)/MgO(2.0\,nm)]$_{10}$ superlattice, computed using the \texttt{nextnano} simulation package \cite{nextnano}. We illustrate the ground and the first-excited state, highlighting their spatial extension across the superlattice. The spin state of the electron is ignored in these calculations.}
    \end{figure}
    
Having established the magnetic order within the sample, we can now address the temperature dependency of the magnetic response. The upper panel of Fig.~\ref{fig:fig4} shows the temperature dependence of the ratio of the remanent and saturation magnetization, $M_{\mathrm{r}}/M_{\mathrm{s}}$, for the sample with $N = 10$. Two distinct temperature regions are observed: between 180 and 330~K, the layers are antiferromagnetically aligned. Below 120~K, the remanence is $M_{\mathrm{r}}/M_{\mathrm{s}} \approx 0.5$, as expected from a $\pi/2$ configuration between adjacent Fe layers. %In addition to this change in magnetic order, a significant change in the saturation field is observed. 
At $T = 300$~K, the saturation field $H_{\mathrm{s}}$ is approximately 14~mT. Upon cooling, $H_{\mathrm{s}}$ increases exponentially, reaching 940~mT at 10~K, as shown in the lower panel of Fig.~\ref{fig:fig4}. A distinct change in the temperature dependence of the coupling strength is observed at 180 K, coinciding with the onset of the reorientation of the magnetic order, from $\pi$ to $\pi/2$. %occurs within the temperature range corresponding to the magnetic reorientation, from $\pi$ to $\pi/2$. 
Notably, the exponential increase in $H_{\mathrm{s}}$ appears to follow similar temperature dependence, above and below the transition region.

Having established the temperature dependence of the saturation field, we now examine the evolution of the switching field of the outermost layers ($H_{\mathrm{1}}$) with respect to the saturation of the sample ($H_{\mathrm{s}}$), which we illustrate in. Fig.~\ref{fig:fig5}. Due to changes in the hysteresis behavior, the temperature range is limited; below 220~K, the discrete hysteresis steps are not observed. Within the accessible range, $H_{\mathrm{1}}$ increases linearly with decreasing temperature, while the saturation field increases exponentially. Consequently, the ratio of the effective coupling of the outermost and innermost layers is found to increase exponentially with decreasing temperature, reaching a value close to 5 at 220~K, as seen in Fig.~\ref{fig:fig5}. 

\begin{figure}[t]
		\centering
		\includegraphics[width=0.95\linewidth]{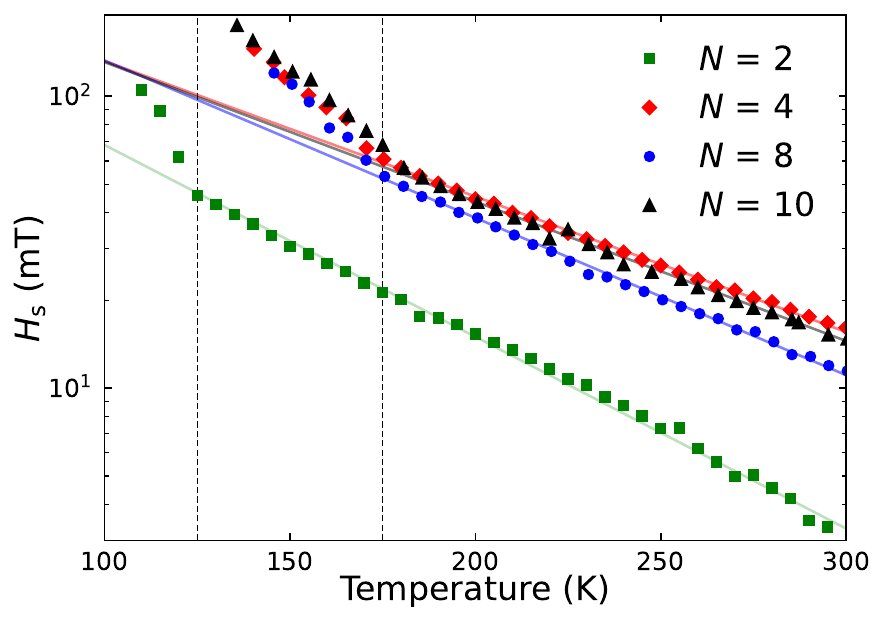}
		\caption{\label{fig:fig8}Temperature dependence of the saturation field ($H_\mathrm{s}$) in samples with $N=$ 2, 4, 8 and 10 bilayer repetitions. Solid lines represent linear fits corresponding to an exponential increase of $H_\mathrm{s}$ with temperature, while dashed lines indicate the temperature below which $H_\mathrm{s}$ deviates from the high temperature behavior.}
    \end{figure}

The temperature dependence of the IEC, along with its sensitivity to the number of bilayer repetitions \( N \), highlights the necessity of incorporating interactions beyond nearest neighbors when describing the magnetic properties of ferromagnet/insulator heterostructures such as Fe/MgO(001). Furthermore, if the tunneling were governed by thermally activated processes \cite{slonczewski_conductance_1989, bruno_theory_1995}, the coupling strength would be expected to increase with temperature—contrary to our experimental observations. We therefore need to ask if our understanding of the interaction in these structures is incomplete. In an attempt to adress that we performed calculations based on a simplified one-dimensional potential well model, not accounting for $e.g.$ spin-dependent effects. The results are presented as an inset in Fig.~\ref{fig:fig5}. The partially delocalized states form \textit{minibands}, with energy level spacings on the order of a few to tens of meV \cite{Handbook_electronic_materials}. In the ground state, the density of states associated with the miniband is lowest near the boundaries of the sample, due to quantum confinement and boundary-induced interference of the wavefunctions. This provides an intuitive rationale for the observed difference in coupling strength between the outermost and the innermost layers. The temperature dependence can also be qualitatively captured by this description: as temperature increases, thermal population of higher miniband states reduces the effective overlap between wavefunctions in adjacent ferromagnetic layers, thereby weakening the interlayer exchange. Notably, this leads to a marked difference in the behavior of the outermost layers in the structure with $N$ = 10 as compared to the one with only two magnetic layers. The outermost layers in the stack of ten has a linear temperature dependence while the coupling strength increases exponentially as temperature decreases for the structure with $N$ = 2, as illustrated in Fig.~\ref{fig:fig8}. As seen in the figure similar trend is observed across all samples. While the saturation field \( H_{\mathrm{s}} \) remains significantly lower for \( N = 2 \) at elevated temperatures, it initially increases more rapidly with decreasing temperature, as compared to the other samples in the series. Assuming the energy levels follow a particle-in-a-box model, the level spacing is expected to scale as \( E_n \approx 1/N^2 \), which is not observed here. Hence, a more detailed description of the interplay of the different length scales is needed to capture the oscillatory dependence on $N$. 

While these simplified analysis provide qualitative insight into the underlying mechanisms, more rigorous theoretical treatments—including spin degrees of freedom—are needed to fully quantify the contributions to the observed temperature dependence. If this interpretation is correct, our results not only advance the understanding of IEC in insulating heterostructures, but also suggest new design principles for achieving functionality in such superlattice-based devices.

\vspace{0.5cm}
\noindent
{\it Data availability}. The data that support the findings of this study are available from the authors upon reasonable request.

\section{Acknowledgments}
The authors acknowledge financial support from the Swedish Research Council (Project No. 2021-00159 and 2019-03581), support with the \texttt{nextnano} simulations and fruitful discussions with Dr. A. Ravensburg and Dr. S. D. Slöetjes. Support by the Super ADAM and ILL staff is also greatly acknowledged.  

%\bibliographystyle{apsrev4-2-titles}
%\bibliography{references}

%

%%%%%%%%%% Merge with supplemental materials %%%%%%%%%%
\pagebreak
\onecolumngrid
\newpage
\begin{center}
\textbf{\large Supplemental Material: Temperature-Induced Magnetic Reorientation and Coupling Enhancement in Fe/MgO(001) Superlattices}
\end{center}
%%%%%%%%%% Merge with supplemental materials %%%%%%%%%%
%%%%%%%%%% Prefix a "S" to all equations, figures, tables and reset the counter %%%%%%%%%%
\setcounter{equation}{0}
\setcounter{figure}{0}
\setcounter{table}{0}
\setcounter{page}{1}
\makeatletter
\renewcommand{\theequation}{S\arabic{equation}}
\renewcommand{\figurename}{Supplementary FIG.}
\renewcommand{\thefigure}{{\bf S\arabic{figure}}}
\renewcommand{\bibnumfmt}[1]{[S#1]}
\renewcommand{\citenumfont}[1]{S#1}
\renewcommand{\thepage}{S-\arabic{page}}
\renewcommand{\thetable}{S-\arabic{table}}
%%%%%%%%%% Prefix a "S" to all equations, figures, tables and reset the counter %%%%%%%%%%

The [Fe/MgO]$_\mathrm{N}$ superlattices were fabricated by magnetron sputtering, with a base pressure of 2$\times$10$^{-9}$~mbar. Sputtering was carried out at an argon working pressure of 2.7$\times$10$^{-3}$~mbar. Prior to deposition, MgO(001) substrates (10$\times$10$\times$1~mm$^{3}$) were annealed at 550~$^\circ$C for 1 hour. During film growth, the substrate temperature was maintained at 165~$^\circ$C. The Fe layers, each with a nominal thickness of $t_{\mathrm{Fe}}$~=~2.0~nm, were deposited by dc sputtering from a high-purity Fe target (99.95~\%), while the MgO layers with a nominal thickness of $t_{\mathrm{MgO}}$~=~1.7~nm were deposited by rf sputtering from a MgO target (99.9~\%). The superlattice structure consisted of $N$ ($N=$ 2--10) Fe/MgO bilayer repetitions, where the first deposited layer was always Fe. The last MgO layer was capped with a 4.2~nm Pd capping layer (dc sputtered) to prevent surface oxidation upon exposure to ambient conditions. The structural properties—namely, layer thickness, interface roughness, and crystallinity—were characterized using x-ray reflectivity (XRR) and x-ray diffraction (XRD), employing Cu K$_{\alpha}$ radiation ($\lambda$ = 0.15418~nm).  A more detailed description of the sample fabrication process and structural characterization can be found in Ref. \cite{warnatz_impact_2021-S}. All the samples were found to be of similar quality as those previously discussed \cite{Moubah2016-S, Magnus2018-S, Anna_Fe_Thickness-S}.

\begin{table}[htbp]
\centering
\renewcommand{\arraystretch}{1.3}
\setlength{\tabcolsep}{15pt}
\caption{Magnetization angles of the Fe layers in a [Fe/MgO]$_{10}$ multilayer at 295 K and 10 K at a field close to remanence, as determined from fits to the PNR data (see Fig.~\ref{fig:fig1} in the main text). An angle of 0$^{\circ}$ corresponds to magnetization aligned parallel to the applied field direction.}
\label{tab:magnetization_angles}
\begin{tabular}{ c c c } 
 \hline
 Fe layer & 295 K ($H_\mathrm{ext}$ = 1.5~mT) &  10 K ($H_\mathrm{ext}$ = 20~mT)  \\ 
 \hline
 1 & 100$^{\circ}$ & 8$^{\circ}$ \\ 
 2 & -73$^{\circ}$ & 61$^{\circ}$ \\
 3 & 106$^{\circ}$ & 13$^{\circ}$ \\
 4 & -72$^{\circ}$ & 65$^{\circ}$ \\
 5 & 100$^{\circ}$ & 4$^{\circ}$ \\
 6 & -82$^{\circ}$ & 72$^{\circ}$ \\
 7 & 92$^{\circ}$ & -1$^{\circ}$ \\
 8 & -83$^{\circ}$ & 74$^{\circ}$ \\
 9 & 100$^{\circ}$ & -14$^{\circ}$ \\
 10 & -90$^{\circ}$ & 68$^{\circ}$ \\ 
 \hline
\end{tabular}
\end{table}
    \begin{figure}[h!]
        \centering
        \includegraphics[width=0.6\linewidth]{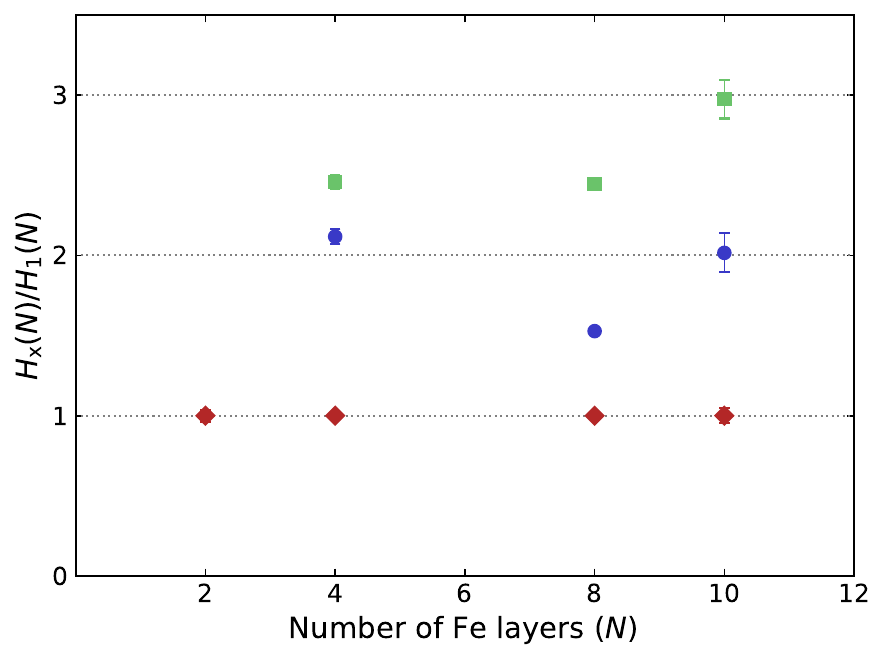}
        \caption{\label{fig:switching_fields}Normalized switching fields. Switching fields $H_{\mathrm{1}}$, $H_{\mathrm{2}}$, and $H_{\mathrm{s}}$ extracted from the room-temperature hysteresis curves shown in Fig.~\ref{fig:fig1}, for $N =$ 2, 4, 8, and 10. Each field is normalized by the corresponding $H_{\mathrm{1}}(N)$ value to highlight variations in relative coupling strength across the different superlattice configurations.}
    \end{figure}
    
The results for all samples are summarized in Fig.~\ref{fig:switching_fields}, where the identified switching fields—$H_{\mathrm{1}}$ (outermost layers), $H_{\mathrm{2}}$ (second switching step), and $H_{\mathrm{s}}$ (saturation field)—are plotted. To highlight the changes in coupling strength with the number of bilayer repeats, each set of switching fields is normalized by $H_{\mathrm{1}}$ for the corresponding sample. A clear change in the relative coupling strength is observed. For instance, in the sample with $N = 10$, the saturation field is approximately three times larger than $H_{\mathrm{1}}$. In the case of nearest neighbor interactions, $H_{\mathrm{s}}$ would be expected to be twice $H_{\mathrm{1}}$. This observation is consistent with magnetic interactions beyond nearest neighbor.

\end{document}